\documentclass[12pt]{article}

\usepackage{color,extarrows} 
\numberwithin{equation}{section}

\newcommand{\be}{\begin{equation}}
\newcommand{\ee}{\end{equation}}
\usepackage{amsmath}
\usepackage{amssymb}
\usepackage{latexsym}
\usepackage{epsfig}
\usepackage{multirow}

\newcommand{\bea}{\begin{eqnarray}}
\newcommand{\eea}{\end{eqnarray}}

\newcommand{\RRR}{{\hbox{\rm R\kern-2.35mm R}}}

\def\ZZZ{{\hbox{ Z\kern-1.6mm Z}}}

\begin{document}

\begin{titlepage}
\rightline{July 2019}
\rightline{  Imperial-TP-2019-CH-05}
\begin{center}
\vskip 2.5cm
{\Large \bf {
Degenerations of K3, Orientifolds and Exotic Branes}}\\
\vskip 2.0cm
{\large {N.  Chaemjumrus and C.M. Hull  }}
\vskip 0.5cm
{\it {The Blackett Laboratory}}\\
{\it {Imperial College London}}\\
{\it {Prince Consort Road}}\\
{\it { London SW7 @AZ, U.K.}}\\

\vskip 2.0cm
{\bf Abstract}
\end{center}

\vskip 0.5cm

\noindent
\begin{narrower}
A recently constructed limit of K3 has a long neck consisting of segments, each  of which is a nilfold fibred over a line,  that are joined together with Kaluza-Klein monopoles. 
The neck is capped at either end by a Tian-Yau space, which is  non-compact,  
hyperk\" ahler and  asymptotic to a nilfold fibred over a line. We show that the type IIA string on this degeneration of K3 is dual to the type I$'$ string, with the Kaluza-Klein monopoles dual to the D8-branes and the Tian-Yau spaces providing a geometric dual to the O8 orientifold planes. At strong coupling, each O8-plane can emit a D8-brane to give an O8$^*$ plane, so that there can be up to 18 D8-branes in the type I$'$ string. In the IIA dual, this phenomenon occurs at weak coupling and there can be up to 18 Kaluza-Klein monopoles in the dual geometry.
We consider further duals in which the Kaluza-Klein monopoles are dualised to NS5-branes or exotic branes.
A 3-torus with $H$-flux can be realised in string theory as an NS5-brane wrapped on $T^3$, 
with the 3-torus fibred over a line.
T-dualising gives a 4-dimensional hyperk\" ahler manifold which is a nilfold fibred over a line, which can be viewed as a Kaluza-Klein monopole wrapped on $T^2$. Further T-dualities then give non-geometric spaces fibred over a line and can be regarded as wrapped exotic branes.
These are all domain wall configurations, dual to the D8-brane. Type I$'$ string theory is the natural home for D8-branes, and we dualise this to find string theory homes for each of these branes. The Kaluza-Klein monopoles arise in the IIA string   on the degenerate K3. 
 T-duals of this give exotic branes on non-geometric spaces.

\end{narrower}

\end{titlepage}

\newpage

\tableofcontents
\baselineskip=16pt
\section{Introduction}

String theory can be defined in 
non-geometric backgrounds 
that are not conventional spacetime manifolds equipped with tensor fields
 --  see~\cite{Hull:2004in,Hellerman:2002ax}, the reviews 
\cite{Wecht:2007wu,Berman:2013eva,Plauschinn:2018wbo} 
and references therein.
  Exotic branes   arise from
 conventional branes after a chain of dualities 
 \cite{Elitzur:1997zn,Hull:1997kb,Blau:1997du,Obers:1998fb}
  and were  associated with 
 non-geometric spaces that are U-folds, with U-duality transition functions, in
  \cite{deBoer:2010ud,deBoer:2012ma}. 
 One of our purposes here is to seek new natural set-ups in string theory that give rise to  consistent configurations of exotic branes. 
 Another related aim is to construct complete consistent   non-geometric string backgrounds that arise from acting on   string vacua with chains of dualities.  We find that  exotic branes naturally live in corresponding non-geometric backgrounds, and that there is a  relation between the brane and the background. Exploring these issues also gives 
 interesting insights into conventional geometric string backgrounds, such as the IIA string on K3.

 The exotic branes that we will consider here can be obtained from D8-branes by a chain of T- and 
 S-dualities, whereas those of~\cite{deBoer:2010ud,deBoer:2012ma} are  dual to D7-branes.
 Type I$'$ string theory provides a consistent string theory home for D8-branes, and has 16 D8 branes moving  on a line interval, with orientifold planes at the ends. The same chain of dualities that takes the 
 D8-branes to the exotic branes takes the type I$'$ string theory to a dual configuration that is the natural home for the exotic branes. 
 One of the dualities in the chain takes the type I$'$ theory compactified on $T^3$ to the type IIA string on K3 while taking the D8-branes to Kaluza-Klein monopoles. It also takes the orientifold planes of the type  I$'$ theory to certain hyperk\" ahler spaces   and gives an interesting picture of strings moving on K3.

 Compactifying type I string theory on a four-torus and T-dualising on all four circles gives an orientifold of IIB string theory.
  The first T-duality of the type I theory takes it to the type I$'$  theory \cite{Polchinski:1995df} with D8-branes between two O8 orientifold planes.
 The next three T-dualities take the D8-branes and O8-planes to D5-branes and O5-planes.
 S-duality takes this to an orbifold of IIB string theory on $T^4$ by $(-1)^{F_L}$ combined with a reflection in the four toroidal coordinates, and takes the D5-branes and O5-planes
 to
 NS5-branes and ON-planes.
 We shall be interested in the result of T-dualising this in 1,2,3 or 4 of the directions that are dual to the original 4-torus.
 From here the first T-duality takes a NS5-brane to a Kaluza-Klein monopole  \cite{Hull:1994ys} and the subsequent T-dualities lead to exotic branes.
 We shall see that the first T-duality leads to Kaluza-Klein monopoles as part of  a geometric background and we shall argue that the subsequent dualities lead to exotic branes in a non-geometric background.

Another motivation for our project is as follows.
The starting point for a much-studied chain of T-dual backgrounds  is the 3-torus with $H$-flux \cite{Kachru:2002sk}.  T-duality on one circle turns off the $H$-flux and gives  a circle bundle over $T^2$ known as a nilfold, 
but further T-dualities give non-geometric spaces. T-dualising the nilfold gives  a T-fold \cite{Hull:2004in}  (a space with T-duality transition functions) and a further T-duality is argued to give a
 non-geometric configuration
  which  is not a conventional space  even locally but can be represented by a doubled geometry \cite{Hull:2009sg}. In \cite{Hull:2019iuy}, these are referred to as {\it essentially doubled} spaces.
In the nomenclature of \cite{Shelton:2005cf} the nilfold has geometric flux or $f$-flux, the T-fold has $Q$-flux and the essentially doubled space has $R$-flux.
However, the 3-torus with flux and its duals do not define worldsheet conformal field theories and so cannot be directly used as string theory backgrounds.

These configurations can arise in string theory, however,  as the fibres   in  string backgrounds which have a  bundle structure, and then these are related by fibrewise dualities, using the adiabatic argumner of~\cite{Vafa:1995gm}.
The simplest case is that in which these spaces are fibred over a line.
There is a hyperk\" ahler metric on the product of the nilfold with a line in which the nilfold moduli depend on the coordinate $\tau $ of the line \cite{Lavrinenko:1997qa,Hull:1998vy,Gibbons:1998ie}.
T-dualising in one direction gives a conformal field theory on  
the product of the line with a 3-torus, which has  constant $H$-flux with a metric and dilaton depending on $\tau$ \cite{Hull:1998vy}.
T-dualising in another direction gives the product of a T-fold 
with a line with the moduli  again depending on $\tau$ \cite{Hull:2004in}.
We aim here to understand these backgrounds  and their implications for string theory better.

The product of the line with a 3-torus with constant $H$-flux can be thought of as an NS5-brane smeared over three directions.
The corresponding supergravity solution can be obtained as follows.
Starting with a D8-brane wrapped on a 3-torus and T-dualising in the three torus directions gives a D5-brane smeared over the   3-torus. The D8-brane supergravity solution~\cite{Bergshoeff:1996ui} is specified by a linear function on the one-dimensional transverse space. T-dualising gives a D5-brane determined by a harmonic function on the  transverse space $\mathbb{R}\times T^3$  which is a linear function on 
 $\mathbb{R}$ and independent of the torus coordinates.
 Then S-dualising gives an NS5-brane smeared over the three torus directions, and this is the product of 6-dimensional Minkowski space with the desired 3-torus with $H$-flux fibred over a line.
 
A  T-duality on a transverse circle takes  the NS5-brane to a Kaluza-Klein monopole \cite{Hull:1994ys}, and so takes the smeared NS5-brane to a smeared Kaluza-Klein monopole, which is  the solution with the nilfold fibred over a line.
A further T-duality on a transverse circle gives what has been termed an exotic brane, referred to in \cite{Obers:1998fb} as a $5_2^2$-brane and in \cite{Hull:1997kb} as a $(5,2^2)$-brane.
The solution with a T-fold fibred over  a line is then seen as a smeared version of the exotic brane.
An interpretation of the unsmeared exotic brane as a T-fold was given in
    \cite{deBoer:2010ud,deBoer:2012ma}.
    A final T-duality gives another exotic brane, which corresponds to the configuration with an essentially doubled space fibred over a line.
    
  T-duality via the Buscher rules \cite{Buscher:1987sk,Buscher:1987qj}  requires an isometry and typically leads to branes smeared over the isometry dimensions, and such smeared branes are often singular.
  However, the smeared branes can be resolved by going to a solution that is localised in the isometric direction, obtained by going to the covering space of an isometric circle, taking a periodic array of localised sources, and then periodically identifying.  
Resolving the smeared supergravity solutions in this way can give interesting string backgrounds and it is these that we will be particularly interested in here.

    The D8-brane is a domain wall separating regions with different values of the Romans mass.
    The various duals of this considered above are then all domain wall solutions too, depending on a single transverse coordinate.
The D8-branes can be consistently incorporated in a  string theory background in the type I$'$ string.
This is the T-dual of the type I string compactified on a circle, and is an orientifold of the type IIA string 
with a vacuum that is the product of a line interval with nine-dimensional Minkowski space. There is an O8 orientifold plane at either end of the interval and 16 D8 branes distributed at arbitrary positions on the interval.
Then following the chain of dualities discussed above
provides a string theory set-up for each of the duals of the torus with flux and the corresponding branes.
These include the NS5-branes with transverse space $T^3\times \mathbb{R}$, Kaluza-Klein monopoles with \lq transverse space' given by a the product of a nilfold with a line, and exotic
$(5,2^2)$-branes with 
\lq transverse space' given by a the product of a T-fold with a line.

We will examine in some detail the   dualities that map the D8-brane wrapped on $T^3$ to 
 the NS5-brane with transverse space given by $\mathbb{R}\times T^3$ and then to a Kaluza-Klein monopole with a Gibbons-Hawking metric which is a circle bundle over a base space
 $\mathbb{R}\times T^2$, giving the product of $\mathbb{R}$ with a   nilfold.
 The D8-branes are domain walls separating regions with different values of the Romans mass, and these are mapped to domain walls separating regions with different values of the $H$-flux on $T^3$ and then to domain walls separating different values of the degree (first Chern class) of the nilfold. The O8 orientifold planes of type I$'$ are mapped to ON5-planes and then to a gravitational version. This gives a picture of KK monopole domain walls distributed along a line interval with some gravitational version of the orientifold planes at either end of the interval.
 
 On the other hand, the same chain of dualities   takes the
 type I$'$ string on $T^3$ to   an orientifold of the type IIB string and then to the  type IIA string theory on K3. Then the K3 geometry and the  KK monopole domain walls distributed along a line give two apparently different representations of what should be the same dual.
  Remarkably, recent work on a  limit of K3 \cite{2018arXiv180709367H} reconciles these two pictures, providing confirmation of our approach. There is a region near the boundary of K3 moduli space in which 
 the K3 develops a long neck which is locally of the form of  the product of a nilfold with a line, with Kaluza-Klein monopoles inserted in that space.
 The ends of the long neck are capped with hyperk\" ahler spaces asymptotic to the product of a nilfold with a line, known as Tian-Yau spaces \cite{Tian1990}. These Tian-Yau caps can be viewed as the duals of the regions around the ON or  orientifold planes and it is remarkable that these are realised as 
smooth geometries, similar to the realisation of certain other duals of  orientifold planes as smooth Atiyah-Hitchin spaces \cite{Sen:1997kz}. The Tian-Yau caps then provide smooth geometries that can be regarded as the gravitational duals of the orientifold planes.
Moreover, the singular domain walls of the supergravity solution obtained by dualising D8-brane solutions are also smoothed out in the K3 geometry as Kaluza-Klein monopole geometries.
The naive duals of the D8 branes would have been smeared Kaluza-Klein monopoles, but in the geometry of  \cite{2018arXiv180709367H} these are resolved to give localised
Kaluza-Klein monopoles.

The classification of Tian-Yau spaces gives the  maximum number of Kaluza-Klein monopoles on the degenerate K3 as 18, which is precisely the maximum number of D8 branes possible in the 
type I$'$ theory \cite{Morrison:1996xf}.
The usual representation of type I$'$ theory has 16 D8 branes and two O8 orientifold planes, but it has been argued that at strong coupling each O8 plane can emit a further D8 brane to leave what has been called an O8$^*$ plane~\cite{Morrison:1996xf,Gorbatov:2001pw}, giving 18 D8 branes and two O8$^*$ planes.
The limiting form of the K3 is the dual of this, with the O8$^*$ planes dualising to Tian-Yau spaces and the D8-branes dualising to KK monopoles.
An O8$^*$ plane with  $n$ coincident D8-branes has charge $n-9$ and dualises to a Tian-Yau space of degree $9-n$. The relation between the IIA string on the degenerate K3 and the type I$'$ string involves an S-duality, so that whereas 18 D8-branes can only be seen at strong coupling in the  type I$'$ string, 18 Kaluza-Klein monopoles can be seen at weak coupling in the type IIA string,

The plan of this paper is as follows. In section 2, we review the nilfold and its T-dualisation to  a 3-torus with $H$-flux, a T-fold and an essentially doubled space.
In section 3, we discuss the string theory solutions obtained by fibring these spaces over a line, to obtain smeared NS5-brane, KK monopole and exotic brane configurations.
In section 4, we the discuss single-sided \lq end of the world' branes that give  supergravity solutions corresponding to orientifold planes.
For the single-sided brane corresponding to the Kaluza-Klein monopole, we discuss the 
 resolution of the singularity with  a Tian-Yau space.  In section 5, we review the type I$'$ theory. In sections 6 and 7, we discuss duals of the type I$'$ theory and their moduli spaces.
 In section 8, we review the explicit construction~\cite{2018arXiv180709367H} of a limit of K3 and discuss how this resolves the singular supergravity solution obtained from dualising the multi-D8 brane configuration arising in   the type I$'$ theory.
 In section 9 we match the moduli spaces of the    type I$'$ theory with those of its duals. In section 10, we extend our discussion to the non-geometric duals involving exotic branes.
 Section 11 presents further discussion.

\section{The Nilfold and its   T-duals }

Consider the 3-torus with $H$-flux given by an integer $m$. The metric and 3-form flux $H$ are
\begin{equation}\label{gH}
ds^2_{T^3}=dx^2+dy^2+dz^2   \qquad  H=mdx\wedge dy\wedge dz
\end{equation}
with  periodic coordinates
$$
x\sim x+2\pi   \qquad  y\sim y+2\pi  \qquad  z \sim  z+2\pi 
$$
Here flux quantisation requires that (in our conventions) $m$ is an integer.

Choosing the 2-form potential $B$  with $H=dB$ 
as
\begin{equation}\label{BB}
B=m\, xdy\wedge dz
\end{equation}
and T-dualising in the $y$ direction gives~\cite{Hull:1998vy} the nilfold ${\cal N}$
\begin{equation}
ds^2_{\cal N}=dx^2+(dy-mxdz)^2+dz^2   \qquad  H=0 \label{nilfold}
\end{equation}
This is a compact manifold that can be constructed as a quotient of the group manifold of the Heisenberg group by a cocompact discrete subgroup; see e.g.~\cite{Hull:2009sg,2018arXiv180709367H}.
It is a circle bundle over a 2-torus, where the 2-torus has   coordinates $x,z$ while the fibre coordinate is $y$. 
Here the first Chern class is represented by $mdx\wedge dz$ and again
$m$, which is the degree of the nilfold,    is required to be an integer.

A complex structure modulus $\tau_1+i \tau_2$  for the 
2-torus with   coordinates $x,z$ and a radius $R$ for the circle fibre can be introduced by choosing the identifications 
$$ (x,z)\sim (x+2\pi ,z), \qquad  (x,z)\sim (x+2\pi \tau_1,z+2\pi \tau_2)
$$
so that for the nilfold we have
\begin{eqnarray}
&(x,y,z)\sim (x+2\pi ,y+2\pi z,z), 
\nonumber \\
& (x,y,z)\sim (x+2\pi \tau_1,y+2\pi \tau_1 z ,z+2\pi \tau_2), 
\nonumber
\\
&(x,y,z)\sim (x,y+2\pi R,z)
\end{eqnarray}
 To simplify our  formulae, we will here display results for the simple case in which $R=1, \tau_1=0,\tau_2=1$; the generalisation of the results presented here to general values of these moduli is straightforward; see e.g. \cite{Hull:1998vy}.

The nilfold can also be viewed as a 2-torus bundle over a circle, with a 2-torus parameterised by $y,z$ and base circle parameterised by $x$. This viewpoint is useful in considering T-duality in the $y$ or $z$ directions, resulting in either case in a fibration over the circle parameterised by $x$ \cite{Hull:2009sg}.

T-dualising the nilfold in the $z$-direction gives
a T-fold ${\cal T}$~\cite{Hull:2004in,Kachru:2002sk,Dabholkar:2002sy}
with
metric and $B$-field given by
\begin{equation}\label{gb}
ds^2_{\text{T-Fold}}=dx^2+\frac{1}{1+(mx)^2}(dy^2+dz^2)   \qquad  B=\frac{mx}{1+(mx)^2}dy\wedge dz 
\end{equation}
which changes by an $O(2,2;\mathbb{Z})$  T-duality transformation under $x\to x+1$, and so
has a T-duality monodromy in the $x$ direction.
A further T-duality in the $x$ direction gives a configuration with explict dependence on a dual coordinate $\tilde x$ and so is not locally geometric but   has a well-defined doubled geometry given in \cite{Hull:2009sg,ReidEdwards:2009nu}. Following \cite{Hull:2019iuy},   we will refer to  such non-geometric spaces with explicit dependence on dual coordinates    as  {\it essentially doubled}.

These examples are instructive  but have  the drawback of not defining a CFT and so not giving a solution of   string theory.
However, these examples can arise in string theory in solutions in which these backgrounds appear as fibres over some base, related by a T-duality acting on the fibres.
The simplest case is that in which these solutions are fibred over a line, defining a solution
that is sometimes referred to as a domain wall background. The cases with a 3-torus or nilfold fibred over a line were  obtained in \cite{Hull:1998vy} from identifications of suitable NS5-brane or KK-monopole solutions, and are dual to D8-brane solutions.
We discuss these and their T-duals in the following section.


\section{Domain walls   from the NS5-brane}
\label{NS5B}
\subsection{The NS5-brane solution}
The NS5-brane supergravity solution  has metric 
\begin{eqnarray}
ds^2_{10} = V(x^i)ds^2(\mathbb{R}^4)+ds^2(\mathbb{R}^{1,5}), \label{NS51}
\end{eqnarray}
where $x^i$ are coordinates of the transverse space  $\mathbb{R}^4$.
(Here $ds^2(\mathbb{R}^d) $, $ds^2(\mathbb{R}^{1,p})$ are the standard flat metrics on
$\mathbb{R}^d$,  $\mathbb{R}^{1,p}$, respectively.)
The function $V(x^i)$ is a harmonic function   satisfying Laplace's equation
\begin{eqnarray}
\nabla^2V(x^i) = 0.
\end{eqnarray}
The $H$-flux   is 
\begin{equation}
H
= *_4dV^{-1}, \label{NS52}
\end{equation}
where $*_4$ is the Hodge dual on the $\mathbb{R}^4$   transverse to the world-volume of the NS5-brane.
Explicitly, this gives
\begin{equation}
H_{ijk}=-\epsilon _{ijkl}\delta ^{lm} \partial _m V
\end{equation}
where $\epsilon _{ijkl}$ is the alternating symbol with $\epsilon _{1234}=1$.
The dilaton is
\begin{equation}
e^{2\Phi} = V. \label{NS53}
\end{equation}
Therefore, the T-duality invariant scalar density $d$ is given by
\begin{eqnarray}
e^{-2d} = e^{-2\Phi}\sqrt{g} = V.
\end{eqnarray}

If $V$ is independent of one or more of the  coordinates of $\mathbb{R}^4$,   the NS5-brane is said to be smeared in those directions. These directions can then be taken to be periodic and we can then T-dualise in them. In what follows, we shall review the various dual spaces that emerge in this way. In each case, we get a string background preserving half the supersymmetry.

We shall be particularly interested in the case in which $V=V(\tau)$ is independent of 3 coordinates, $x,y,z$, which can then all be taken to be periodic. The NS5-brane is then  smeared over the $x,y,z$ directions, and   there is $H$-flux on the 3-torus   with coordinates $x,y,z$. The solution (\ref{NS51}), (\ref{NS52}), and (\ref{NS53}) then represents the product of flat 6-dimensional space $\mathbb{R}^{1,5}$ with the 4-dimensional space given by 
a 3-torus with $H$-flux fibred over a   line with coordinate $\tau$~\cite{Hull:1998vy}.
Successive T-dualities will then take the 3-torus with flux fibred over a line first to a nilfold fibred over a line, then  to a T-fold fibred over a line and finally to an  essentially  doubled  space (with $R$-flux)   fibred over a line.

Consider first the case in which
 $V$ is independent of one of the coordinates $x^i=(\tau,x,y,z)$, $y$ say. We  take $y$ to be periodic (i.e. we can identify under $y\to y+2\pi $) and T-dualise in the $y$ direction to obtain the KK-monopole solution
\begin{eqnarray}
ds^2_{10} =  ds^2(GH)+ds^2(\mathbb{R}^{1,5}), \label{KKsolution}
\end{eqnarray}
where $ ds^2(GH)$ is a Gibbons-Hawking metric 
\begin{eqnarray}
 ds^2(GH)=V(d\tau^2 +dx^2+ dz^2) + V^{-1}(dy+\omega)^2 \label{metricGH}
 \end{eqnarray}
with $V(\tau,x,z)$  a harmonic function on $\mathbb{R}^3$ and
$\omega $ a 
1-form on $\mathbb{R}^3$ 
satisfying
\begin{eqnarray}
\vec{\nabla}\times\vec{\omega} = \vec{\nabla}V
\end{eqnarray}
 and 
\begin{equation}
H=0, \qquad \Phi= \rm{constant}.
\end{equation}
The   metric   is hyperk\" ahler.

If $V$ is   independent of $y,z$, so that $V=V(\tau,x)$ is a function on $\mathbb{R}^2$, then the smeared NS5-brane solution can be taken to be periodic in the $y,z$ directions and can be dualised in one or both directions.
T-dualising in both the $y$ and $z$ directions is the same as T-dualising the KK-monopole solution (\ref{KKsolution}) in the $z$ direction and gives 
\begin{equation}
ds^2= ds^2(X)+ds^2(\mathbb{R}^{1,5}),
\end{equation}
where $X$ is a four-dimensional space with metric
\begin{eqnarray}
ds^2(X) = V(d\tau^2+dx^2)+\frac{V}{V^2+w^2}(dy^2+dz^2) \label{metricX}
\end{eqnarray}
with $V(\tau,x)$ a harmonic function on $\mathbb{R}^2$ and $w$   a function on $\mathbb{R}^2$ which satisfies
\begin{eqnarray}
\frac{\partial V}{\partial \tau} = \frac{\partial w}{\partial x}, \qquad \frac{\partial V}{\partial x} = -\frac{\partial w}{\partial \tau}\label{w}
\end{eqnarray}
so that $w$ is also harmonic, and
\begin{eqnarray}
B = \frac{w}{V^2+w^2}dy\wedge dz.
\end{eqnarray}
To find the dilaton, we note that
\begin{eqnarray}
e^{-2d} \equiv e^{-2\Phi}\sqrt{g}
\end{eqnarray}
is invariant under T-duality, so that if under T-duality
\begin{eqnarray}
g_{\mu \nu }\to g' _{\mu \nu }, \qquad  \Phi\to \Phi'
\end{eqnarray}
we have that 
\begin{eqnarray}
\Phi'= \Phi + \frac 1 2  \log (\sqrt{g'}/\sqrt{g}).
\end{eqnarray}
For the metric (\ref{metricGH}) we have $\sqrt{g}=V$ 
and take $\Phi = \Phi_0$ for some constant $\Phi_0$,
while for (\ref{metricX}) we have
\begin{eqnarray}
\sqrt{g'} = \frac{V^2}{V^2+w^2}\, ,
\end{eqnarray}
so the dilaton for the space $X$ is 
\begin{eqnarray}
\Phi =\Phi_0+  \frac{1}{2}\log\left(\frac{V}{V^2+w^2}\right).
\end{eqnarray}

Finally, if $V$ is   independent of $x,y,z$, so that $V=V(\tau)$, then we can take $x,y,z$ as periodic and can T-dualise in one, two or three directions.
For $V=V(\tau)$ to be a harmonic function on $\mathbb{R}$, $V''=0$, it must be a linear function. The simplest case is to take
\begin{equation}
V(\tau) = m\tau+c,
\end{equation}
where $m$ and $c$ are constant. The form $V(\tau)$ implies the form of $w$ (\ref{w}) as
\begin{eqnarray}
w = V'(\tau)x = mx.
\end{eqnarray}
Then for the NS5-brane solution we obtain the following conformally flat metric on 
$T^3\times \mathbb{R}$
\begin{equation}
ds^2 = V(\tau)[d\tau^2+dx^2+dy^2+dz^2]
\label{NS1}
\end{equation}
together with the   $H$-flux  on $T^3$
\begin{eqnarray}
H = *_4dV^{-1} = -mdx\wedge dy\wedge dz \label{NS2}
\end{eqnarray}
and dilaton
\begin{equation}
e^{2\Phi} = V(\tau) \label{NS3}.
\end{equation}
The usual flux quantisation condition implies that the coefficient $m$ is quantized; we adopt conventions in which $m$ is an integer.
By changing coordinates $\tau \to \sigma(\tau) = \log V$, one could arrange for the dilaton to have linear dependence on the coordinate $\sigma$.
This 4-dimensional space has topology $\mathbb{R}\times T^3$. The geometry is a 3-torus with flux fibred over $\mathbb{R}$ -- the metric of the 3-torus and the dilaton depend on the coordinate $\tau$, but 
the  flux  $\int _{T^3} H=m $ remains constant and is quantised.

More generally, $V$ can be taken to 
be piecewise linear,
e.g.
\begin{equation}
V(\tau) = \begin{cases}
c+m'\tau, & \tau \le 0 \\
c+m\tau , & \tau>0.
\end{cases}\label{Vtau}
\end{equation}
This is continuous but not differentiable at $\tau =0$. The singularity at $\tau =0$ corresponds to a domain wall  at $\tau =0$ separating two \lq phases' with fluxes $m,m'$. This can be thought of as a brane that has a tension proportional to $m-m'$.
The solution can be understood  \cite{Hull:1998vy} as a dual of  the D8-brane solution \cite{Bergshoeff:1996ui}, as we will discuss in section \ref{secD8}.  A full string solution is then obtained by introducing O8-planes in the D8-brane solution and dualising.

A multi-brane solution with  domain walls at $\tau = \tau_1,\tau_2,\dots \tau_n$ is given by
\begin{equation}
V(\tau) = \begin{cases}
c_1+m_1\tau, & \tau \le \tau_1 \\
c_2+m_2\tau, & \tau_1 < \tau \le \tau_2\\
\vdots & \\
c_n+m_n\tau, & \tau_{n-1} < \tau \le \tau_n\\
c_{n+1}+m_{n+1}\tau, & \tau > \tau_n
\end{cases}\label{Vtau1taun}
\end{equation}
for some constants $c_1, m_i$,
and for continuity the constants $c_r$ for $r>1$ are given in terms of $c_1, m_i$ by 
\begin{equation}
c_{r+1}=c_r + (m_r-m_{r+1})\tau _r.
\end{equation}
The brane charge of the domain wall at $\tau _r$ is the integer
\begin{equation}
N_r=m_{r+1}- m_r .
\end{equation}
Note that the derivative $M(\tau)\equiv V'(\tau)$ of $V(\tau)$ with respect to $\tau$  is piece-wise constant away from the domain wall points $\tau _r$:
\begin{equation}
M(\tau) = \begin{cases}
m_1, & \tau < \tau_1 \\
m_2, & \tau_1 < \tau < \tau_2\\
\vdots & \\
m_n,& \tau_{n-1} < \tau < \tau_n\\
m_{n+1},& \tau > \tau_n.
\end{cases}\label{M}
\end{equation}
The solution is then given by the metric (\ref{NS51}) and dilaton (\ref{NS53}) with (\ref{Vtau1taun})
and  the   $H$-flux  on $T^3$ given by
\begin{eqnarray}
H = *_4dV^{-1} = -Mdx\wedge dy\wedge dz. \label{NS2}
\end{eqnarray}

Taking the product of the solution (\ref{NS1}), (\ref{NS2}) with 6-dimensional Minkowski space $\mathbb{R}^{1,5}$ gives a space 
$\mathbb{R}\times T^3\times \mathbb{R}^{1,5}$ with NS5-branes (smeared over the $T^3$) inserted at $\tau=\tau _i$.
The transverse space for the NS5-branes is
$\mathbb{R}\times T^3$.

\subsection{Nilfold   background}
The background (\ref{NS1}) has isometries in the  $x, y$ and $z$ directions. Performing T-duality in the $y$-direction   gives a background that is $\mathbb{R}\times {\cal N}$,
with a metric dependent on the coordinate $\tau $ of $\mathbb{R}$, so that it is
nilfold fibred over the real line.
 The metric is
\begin{equation} 
ds^2 =V(\tau)(d\tau^2+dx^2+ dz^2 ) + \frac{1}{V(\tau)}(dy+M(\tau)xdz)^2 \label {nilfold1}
\end{equation}
 and the other fields are trivial:
\begin{equation}
H=0, \qquad \Phi= \rm{constant}. \label{nilfold2}
\end{equation}
This  4-dimensional metric   can be viewed as a Gibbons-Hawking metric with a harmonic function $V$ depending linearly on a single coordinate. It preserves half the supersymmetry and so 
 is hyperk\" ahler.
 The three complex   structures are given by
\begin{eqnarray}
J^1 &=& d\tau\wedge (dy+M(\tau)xdz)+V(\tau)dx\wedge dz\\
J^2 &=& dx\wedge(dy+M(\tau)xdz)+V(\tau)d\tau\wedge dz\\
J^3 &=& dz\wedge(dy+M(\tau)xdz)+V(\tau)d\tau\wedge dx.
\end{eqnarray}
For $V$ of the form (\ref{Vtau1taun}) we have a multi-domain wall solution and we will refer to this as the hyperk\" ahler wall solution $\hat {\cal N}$. In the region between walls $\tau _i<\tau <\tau _{i+1}$ or for $0<\tau <\tau _{1}$ or $\tau _n<\tau <\pi$
it has the topology 
$I\times {\cal N}$ where $I$ is a line interval. The dilaton is constant and the antisymmetric tensor gauge 
field strength $H=dB$ is zero, but the metric depends on the coordinate $\tau$.
The space is singular at the end points $\tau = 0$, $\tau = \pi$ 
and the domain wall positions $\tau =  \tau _i$;
 we will discuss the resolution of these singularities in later sections.

For fixed $\tau$, the geometry is a nilfold, which can be viewed as a circle bundle over a 2-torus, where the circle fibre has coordinate $y$ and the torus base has coordinates $x,z$.
The geometry is warped by the factor of $V$: the
circumference of the circle fibre is $1/V$ while the circumference of each of the circles of the 2-torus is $V$.

Taking the product 
${\cal {N}}
\times
 \mathbb{R}
 ^{1,5}$ 
of the hyperk\" ahler wall solution with 6-dimensional Minkowski space gives a space which can be viewed as a background with  Kaluza-Klein monopoles.
The  usual Kaluza-Klein monopole is given by the product of self-dual Taub-NUT space with $\mathbb{R}^{1,5}$. This 
 has a
 transverse space that  is $\mathbb{R}^{3}$,
  and the Taub-NUT space is a circle bundle over this space (with a point removed).
 We shall be interested later in the Kaluza-Klein monopole  whose  transverse space  is $\mathbb{R}\times T^2$, 
 so that the Gibbons-Hawking circle bundle over the transverse space is $\mathbb{R}\times {\cal {N}}$.
Here, we are obtaining a version of this smeared over the coordinates of the transverse $T^2$,
 and the geometry  can be thought of as $\mathbb{R}\times {\cal {N}}\times \mathbb{R}^{1,5}$ with a  Kaluza-Klein monopole at each $\tau _i$, smeared over the $T^2$ fibres.

\subsection{T-fold and R-fold backgrounds}

Performing T-duality along the $z$-direction results in the T-fold ${\cal T}$ fibred over a line
\cite{Hull:2004in,Ellwood2006}.
 The metric and B-field of this background are given by 
\begin{equation}
ds^2 = V(\tau)(d\tau)^2+V(\tau)(dx)^2+\frac{V(\tau)}{V(\tau)^2+(M(\tau)x)^2}(dy^2+dz^2), \label{Tfold1}
\end{equation}

\begin{equation}
B = \frac{M(\tau)x}{V(\tau)^2+(M(\tau)x)^2} dy\wedge dz \label{Tfold2}
\end{equation}
while  the dilaton is
\begin{equation}
  \Phi=  \frac{1}{2}\log\left(\frac{V(\tau)}{V(\tau)^2+(M(\tau)x)^2}\right). \label{Tfold3}
  \end{equation}
For fixed $\tau$ we obtain the T-fold (\ref{gb}).
In the region between walls $\tau _i<\tau <\tau _{i+1}$ the space is the product of the interval $I$ with the T-fold (\ref{gb}) with the fields depending on $\tau$ through the 
 warp by factor  $V$.

Finally, a further T-duality in the $x$ direction gives an essentially doubled space which is not locally geometric but which has a well-defined doubled geometry which is given  by the doubled configurations of  \cite{Hull:2009sg,ReidEdwards:2009nu} fibred over a line. This will be discussed further elsewhere.

\subsection{Exotic Branes and T-folds}

T-dualising the NS5-brane on a transverse circle gives a Kaluza-Klein monopole \cite{Hull:1994ys}.
The transverse space of the NS5 brane is taken to be $S^1\times \mathbb{R}^3$ and the harmonic function determining the solution is taken to be independent of the circle coordinate, so 
that the NS5-brane is smeared over the transverse circle. Then both the KK-monopole and the smeared NS5 solutions are given by a harmonic function on $\mathbb{R}^3$.

If the NS5-brane solution is smeared over $T^2$, then a further T-duality is possible.
The transverse space of the NS5 brane is now  taken to be $T^2 \times \mathbb{R}^2$ and the solution is given by a harmonic function on $\mathbb{R}^2$.
T-dualising on one circle gives a KK-monopole smeared over a circle with transverse space
$\mathbb{R}^2\times S^1$ and  the solution  is given by a harmonic function on $\mathbb{R}^2$.
The harmonic function in $\mathbb{R}^2$ leads to a monodromy round each source.
A further T-duality on a transverse circle gives what has been termed an exotic brane, referred to in \cite{Obers:1998fb} as a $5_2^2$-brane and in \cite{Hull:1997kb} as a $(5,2^2)$-brane.
This was interpreted in   \cite{deBoer:2010ud,deBoer:2012ma}
as a T-fold, with a T-duality monodromy round each source in the transverse $\mathbb{R}^2$.

Here we are interested in an NS5-brane smeared over $T^3$, so that the transverse space is $T^3 \times \mathbb{R}$ and the harmonic function is a linear function on $\mathbb{R}$.
Then the first T-duality gives a KK-monopole smeared over $T^2$ with transverse space $T^2 \times \mathbb{R}$. The second T-duality gives a
an exotic $5_2^2$-brane or $(5,2^2)$-brane smeared over $S^1$
with transverse space $S^1 \times \mathbb{R}$. 
A third T-duality gives an exotic brane referred to in \cite{Hull:1997kb} as a $(5,3^2)$-brane.  In the notation of \cite{Obers:1998fb}, this would be a $5_2^3$-brane.
This is an essentially doubled solution with $R$-flux.

The background (\ref{NS1}), (\ref{NS3}), (\ref{NS2}) was interpreted as $\mathbb{R}\times T^3\times \mathbb{R}^{1,5}$ with smeared NS5-branes    inserted at $\tau=\tau _i$ and the 
geometry (\ref{nilfold1}), (\ref{nilfold2})
was
thought of as $\mathbb{R}\times {\cal {N}}\times \mathbb{R}^{1,5}$ with a smeared Kaluza-Klein monopole at each $\tau _i$.
In the same spirit, the T-fold solution (\ref{Tfold1}), (\ref{Tfold2}), (\ref{Tfold3}) can be thought of as the T-fold background
$\mathbb{R}\times {\cal {T}}\times \mathbb{R}^{1,5}$ with a smeared exotic brane at each $\tau _i$.

For the  T-duality of the NS5-brane on a transverse circle, it is not necessary to assume that the NS5-brane is smeared over the circle. 
Instead, one can take 
an NS5-brane localised on $S^1\times \mathbb{R}^3$. This can be constructed by taking a periodic array of  NS5-branes on $\mathbb{R}^4$
located at points arranged on a line $(0,0,0, 2\pi R m)$ for integers  $m=0,\pm1,\pm 2,\dots$ and then identifying $x^4\sim x^4+2\pi R$. 
Summing the contributions from the sources as in~\cite{Ooguri:1996me}
gives a 
 harmonic function  defining the  NS5-brane solution that depends explicitly on $x^4$. The T-duality of this solution has been discussed in \cite{Gregory:1997te}; see also
\cite{Tong:2002rq,Harvey:2005ab,Dabholkar:2005ve,Jensen:2011jna}.
The T-dual solution is essentially doubled, depending explicitly on the coordinate $\tilde x^4$ of the T-dual  circle.
This then gives a modification of the Kaluza-Klein monopole geometry with explicit dependence on the dual coordinate.

\section{Single-sided  domain walls and   Tian-Yau Spaces} \label{TYsec}

Consider a domain wall of the kind discussed in the previous subsections, with a profile given by a function $V(\tau)$ of the transverse coordinate.
Taking
 \begin{equation}
\label{vissts}
V= c+ m|\tau |
\end{equation}
gives a domain wall at $\tau =0 $ of charge $m$ that is invariant under the reflection
\begin{equation}
\tau \to - \tau.
\end{equation}
Quotienting by this reflection identifies  the half-line $\tau <0$ with the half-line $\tau >0$ resulting in a single-sided domain wall solution defined for $\tau \ge 0$, with a singular wall at $\tau =0$
\cite{Gibbons:1998ie}.
For the applications to string theory in the next section, we will be interested in 
the case in which the $\tau$ direction is a line interval with a single-sided domain wall at either end.

For the case of the nilfold fibred over a line with metric (\ref{nilfold1}) with $V$ given by (\ref{vissts}), 
this orbifold singularity has a remarkable resolution to give a smooth manifold,
as was proposed in \cite{Gibbons:1998ie}. The Tian-Yau space \cite{Tian1990} is a smooth four-dimensional  hyperk\" ahler manifold
fibred over the half-line $\tau >0$ such that for large $\tau$ it approaches the metric (\ref{nilfold1}) for a nilfold fibred over a line, so that it can be regarded as a resolution of the single sided brane \cite{2018arXiv180709367H}.

The Tian-Yau space \cite{Tian1990} is a complete non-singular non-compact hyperk\" ahler space that is asymptotic to a nilfold bundle over a line.
It is 
of the form $M\setminus D$, where $M$ is a del Pezzo surface and $D \subset M$ is a certain
$T^2$ submanifold
 (which is a smooth anti-canonical divisor).  
The del Pezzo surfaces are complex surfaces   classified by their degree $b$, where $b = 1, 2, \ldots, 9$. The  del Pezzo surface of degree nine is the complex projective space $\mathbb{CP}^2$. A del Pezzo surface $M_b$ of degree $b$ can be constructed by blowing up a point in the del Pezzo surface of degree $b+1$, $M_{b+1}$. 
That is, a degree $b$ del Pezzo surface can be constructed from blowing up $9-b$ points in $\mathbb{CP}^2$, and there are restrictions on the positions of the points that can be blown up.
There are two types of del Pezzo surface of degree eight,  which are $\mathbb{CP}^1\times\mathbb{CP}^1$ and the result of blowing up one point in $\mathbb{CP}^2$.

The Tian-Yau space $X_b = M_b \backslash D$ is a non-compact space that is asymptotic to   $\mathbb{R}\times \mathcal{N}_b$, where $\mathcal{N}_b$ is a nilfold
   of  degree $m = b$. 
In the asymptotic region, the Tian-Yau metric can be approximated by the metric (\ref{nilfold1})
 where $V(\tau)$ is a non-zero linear function $V= c+ m\tau $ so that  $V(\tau) \rightarrow \infty$ as $\tau \rightarrow \infty$.
The degree $m$ of the nilfold is given by the degree $b$ of the del Pezzo surface, so  only degrees $m = 1, 2, \ldots, 9$ can arise.

   Starting from a del Pezzo surface  of degree nine and blowing up nine points gives  a rational elliptic surface $M$ \cite{Hein2012}. It will be convenient to refer to  this case as a Tian-Yau space of zero degree, 
so that we can  extend the range of the degree to $b = 0,1, 2, \ldots, 9$.
This   $b=0$ zero degree space has the structure of an elliptic fibration, $f:M \rightarrow \mathbb{CP}^1$ with the fiber being an elliptic curve or 2-torus. For zero degree, the nilfold reduces to  a 3-torus and the space $M \backslash D$ is  a non compact space that is asymptotic to a cylinder $T^3\times I$. It is then an ALH gravitational instanton. 

The $m\ne 0$ case gives a generalisation of the ALH case which is asymptotic to the product of a nilfold of degree $m$ with an interval. 
Changing variables to 
$s= (m\tau )^{3/2}$, the metric takes the following asymptotic form for large $s$:
\begin{equation}
g\sim \frac 4 {9m^2} ds^2 + s^{2/3}(dx^2+dz^2) +s^{-2/3}
(dy+bxdz)^2.
\end{equation}
Then for large $s$, the size of the $S^1$ fibres falls off as $s^{-1/3}$ while the size of each 1-cycle of the 2-torus base grows as $s^{1/3}$.

\section{The D8-Brane and Type I$'$ String Theory} \label{secD8}

\subsection{The D8-Brane and its Duals}

The D8-brane solution of  the IIA string \cite{Bergshoeff:1996ui} has string-frame metric 
\begin{equation}
ds^2=V^{-1/2} 
ds^2(\mathbb{R}^{1,8})
 +V^{1/2}d\tau ^2\label{metrictypeI'}
\end{equation}
with dilaton
\begin{equation}
e^\Phi = V(\tau)^{-\frac{5}{4}} \label{phitypeI'}
\end{equation}
and RR field strength
\begin{eqnarray}
F_{(10)} &=& dt\wedge dx_1\wedge \cdots \wedge dx_8 \wedge d(V(\tau)^{-1}) \nonumber\\
&=&  -\frac{M(\tau)}{V^2(\tau)}dt\wedge dx_1\wedge \cdots \wedge dx_8 \wedge d\tau \nonumber\\
&=& -M(\tau) \Omega _{\text{Vol}}\label{FtypeI'}
\end{eqnarray}
where
\begin{equation}
 \Omega _{\text{Vol}} = \sqrt{-g}dt\wedge dx_1\wedge \cdots \wedge dx_8 \wedge d\tau
\end{equation}
is the volume form.
The Hodge dual of $F_{(10)}$ is a zero-form,
\begin{eqnarray}
F_{(0)} = -M(\tau).
\end{eqnarray}
This zero-form $F_{(0)}$ gives the  mass parameter in the massive type IIA supergravity \cite{Romans:1985tz}. For our solution it is piece-wise constant as in (\ref{M}), so that the Romans mass parameter is different on either side of a domain wall.
There is an 8+1 dimensional longitudinal space and a one-dimensional transverse
space with
coordinate
$\tau$. 
Here $V(\tau)$ is piecewise linear. Taking it to be of the form (\ref{Vtau}) gives a D8-brane of charge $m-m'$ 
at 
$\tau=0$ while the multi-brane solution (\ref{Vtau1taun}) is a  multi-brane solution with D8-branes at positions $\tau_1, \cdots, \tau_n$.

If three of the transverse dimensions are compactified to a 3-torus with coordinates $x,y,z$,
 the metric can be written as
 \begin{equation}
ds^2=V^{-1/2} 
[dx^2+dy^2+dz^2+ ds^2(\mathbb{R}^{1,5})]
 +V^{1/2}d\tau ^2 \, .
\end{equation}
T-dualising in the $x,y,z$ directions gives a D5-brane IIB solution smeared over   $T^3$
\begin{equation}
ds^2=V^{-1/2} 
ds^2(\mathbb{R}^{1,5})
 +V^{1/2}(d\tau ^2+dx^2+dy^2+dz^2)\label{metricIIB}
\end{equation}
\begin{equation}
F_{(7)}= dt\wedge dx_1 \wedge \cdots \wedge dx_5 \wedge d(V(\tau)^{-1}) \label{FIIB}
\end{equation}
\begin{equation}
e^{2\Phi} = V(\tau)^{-1} \label{phiIIB}
\end{equation}
where $V$ is of the form (\ref{Vtau}), or (\ref{Vtau1taun}).
The Hodge dual of $F_{(7)}$ is the RR field strength three-form, which is given by
\begin{equation}
F_{(3)} = *F_{(7)} = -M(\tau)dx\wedge dy\wedge dz.
\end{equation}
Next, S-duality gives a smeared NS5-brane solution, 
which is precisely the solution (\ref{NS1}), (\ref{NS3}), and (\ref{NS2}) given in section \ref{NS5B},  with a transverse space given by a $T^3$ bundle over a line.
We will discuss further T-duals of this background in sections \ref{modsp} and \ref{ngdu}.

\subsection{The type I$'$ String}

The multi-D8 brane solution has a dilaton depending   on the   transverse coordinate in such a way that the  dilaton becomes large and hence the string becomes strongly coupled in certain regions.
A well-behaved solution of string theory with D8-branes arises in the type I$'$ string \cite{Polchinski:1995df}, which arises from
compactifying the type I string on a circle and T-dualising. This can be viewed as an orientifold of the type IIA string compactified on a circle, resulting 
in a theory on $S^1/\mathbb{Z}_2$ with O8 orientifold planes introduced at the fixed points, and with 16 D8-branes (together with their mirror images under the action of $ \mathbb{Z}_2$) located at arbitrary locations. It can also be viewed as a theory on the interval $I$ arising from the quotient $S^1/\mathbb{Z}_2$  with O8-planes at the end points, and with 16 D8-branes located at arbitrary points on the interval.

The supergravity solution corresponding to the   type I$'$ string with 16 D8-branes has string frame metric
\begin{equation}
ds^2=V^{-1/2} 
ds^2(\mathbb{R}^{1,8}) 
 +V^{1/2}d\tau ^2, 
\end{equation}
where $V(\tau)$ is a harmonic function on 
the interval $I$ with coordinate $\tau$, $0\le \tau \le \pi$.
The dilaton  is given by
\begin{eqnarray}
e^{\Phi} = V(\tau)^{-\frac{5}{4}}
\end{eqnarray}
and the RR field strength is given by
\begin{eqnarray}
F_{(0)} = -M(\tau)
\end{eqnarray}
where $M(\tau)\equiv V'(\tau)$.

The orientifold planes are located at $\tau = 0$ and $\tau = \pi$, while the   16 D8-branes are located at arbitrary points, $\tau_1, \ldots, \tau_{16}$ between $\tau = 0$ and $\tau = \pi$. 
The function $V(\tau)$ is piecewise linear and,
for general positions of the D8-branes, it is given by (\ref{M})  with $n=16$, $m_1=-8$, $m_{i+1}=m_i +1$ so that $m_i=i-9$ and $m_{17}=8$, so that the orientifold planes are treated as
sources of charge $-8$ at the end-points of $I$.
The gradient of $V$ jumps by $+1$ at each D8-brane.
 Then
 \begin{equation}
V(\tau) = \begin{cases}
c_1-8\tau, & 0\le \tau \le \tau_1 \\
c_2-7\tau, & \tau_1 < \tau \le \tau_2\\
\vdots & \\
c_i + (i-9)\tau, & \tau_{i-1} < \tau \le \tau_i\\
\vdots & \\
c_{16}+7\tau, & \tau_{15} < \tau \le \tau_{16}\\
c_{17}+8\tau, & \tau_{16} < \tau \le \pi
\end{cases}
\end{equation}
with 
\begin{equation}
c_{r+1}=c_r -\tau _r.
\end{equation}
The slope of the function $V$ is $M(\tau)\equiv V'(\tau)$ and is given by 
\begin{equation}
M(\tau) = \begin{cases}
-8, & 0\le  \tau < \tau_1 \\
-7, & \tau_1 < \tau < \tau_2\\
\vdots & \\
7,& \tau_{15} < \tau \le \tau_{16}\\
8,&
 \tau_{16} < \tau \le \pi
.
\end{cases}
\end{equation}

For $r$ coincident branes with $\tau_i =\tau_{i+1}=\dots =\tau _{i+r-1}$ the slope jumps from 
 $m_i $ for $\tau_{i-1}< \tau <  \tau_i $ to $m_i +r$ for $\tau_{i}< \tau <  \tau_{i+1} $.
In general this leads to domain walls at $\tau = \tau_1,\tau_2,\dots \tau_n$
with positive charges
$N_1, N_2,\dots N_n $
corresponding to $N_i$ D8-branes at $\tau_i$.
If there are $N_0$ D8-branes coincident with  the O8-plane at $\tau=0$ and 
$N_{n+1}$ D8-branes coincident with  the O8-plane at $\tau=\pi$
then
\begin{equation}
\sum _{i=0} ^ {n+1} N_i =16.
\end{equation}
Then $V$ is given by (\ref{Vtau1taun}) with $N_r=m_{r+1}- m_r $
and
\begin{equation}
m_1=-(8-N_0), \qquad m_{n+1}=8-N_{n+1}
\end{equation}
while $M$ is (\ref{M}).

At strong coupling, new effects can arise. Each orientifold plane can emit a further D8-brane at strong coupling, so that an O8 plane of charge $-8$ emits a D8-brane of charge $+1$ to leave an 
O8$^*$ plane of charge $-9$~\cite{Morrison:1996xf,Gorbatov:2001pw}.
While at weak coupling there are 16 D8  branes, at strong coupling   there can be 17 or 18 D8 branes; this
leads to  a enhancement of the gauge symmetry, e.g. to $SU(18)$ that is non-perturbative in the type I$'$ theory
\cite{Morrison:1996xf,Gorbatov:2001pw}.

\section{Duals and Moduli Spaces} \label{dualsect}

\subsection{Duals of the type I$'$ String}  

The type I string can be obtained as an orientifold of the type IIB string
\begin{equation}
\text{I}= \frac {\text{IIB} }{ \Omega}
\end{equation}
where $\Omega$ is the world-sheet parity operator.
This has 16 D9-branes to cancel the charge of the O9-plane.
Compactifying on a circle in the $X^9$ direction and T-dualising gives the type I$'$ string.
This is now a quotient of the IIA string \cite{Polchinski:1995df}
\begin{equation}
\text{I}'= \frac {\text{IIA} }{ \Omega R_9} \label{typeI'}
\end{equation}
where $R_9$ is   reflection in  $X^9$.
This results from the fact that the  T-duality $T_i$ in the $X^i$ direction acts as
\cite{Polchinski:1996fm}
\begin{equation}
T_i: \Omega \to \Omega R_i.
\end{equation}
The periodic coordinate $X^9\sim X^9+ 2\pi$ is identified under the action of the reflection $R_9:X^9\to -X^9$ so that after the quotient, the $X^9$ circle becomes $S^1/\mathbb{Z}_2$. An orientifold O8-plane is introduced at each of the two  fixed points, $X^9=0$
and $X^9=\pi$. The O8 planes each have charge $-8$ (in units in which a single D8-brane has charge $+1$) and this is cancelled by the charge of the 16 D8-branes 
arising from the T-dual of the 16 D9-branes of  the type I string.
The space $S^1/\mathbb{Z}_2$ can be viewed as the interval $0\le X^9\le \pi$, with an orientifold plane at either end of the interval.

Next, compactifying on $X^8$ and T-dualising might be expected to give, from (\ref{typeI'}), an orientifold of the IIB string by $ \Omega R_{89}$ where 
$R_{89}=R_8R_9$. However, this leads to a problem, as this orientifold is not supersymmetric, but T-duality is expected to preserve supersymmetry.
The resolution of this  \cite{Dabholkar:1996pc}  results from the fact that the T-dualities $T_8$ and $T_9$ do not commute in the superstring. The T-duality $T_i$ in the $X^i$ direction
acts as a reflection on the left-moving bosonic world-sheet fields: 
\begin{equation}
(X_L^i,X_R^i)
\to
(-X_L^i,X_R^i).
\end{equation}
On  the left-moving spin-fields $S_L$, this reflection acts through $
S_L\to t_i S_L$ where  $t_i =\Gamma ^{11}\Gamma_i$.
As $t_i$ and $t_j$ anticommute for $i\ne j$, it follows that
$T_iT_j\ne T_jT_i
$ when acting on fermions, 
 the result of two T-dualities is only determined up to a factor of $(-1)^{F_L}$.
This ambiguity gives two possible ways of  taking two T-duals of the type I string: one way  gives IIB$/\Omega R_{89}$ which is not supersymmetric, and the other gives IIB$/\Omega R_{89}(-1)^{F_L}$ which is supersymmetric. Here and in each case that follows, we will define T-duality to be the transformation that preserves supersymmetry. See \cite{Dabholkar:1996pc,Dabholkar:1997zd,Hanany:2000fq}
for further discussion.

In this way, we obtain the standard chain of supersymmetric orientifolds by successive T-dualities for the type I string compactified on the 4-torus in the $ X^6,X^7,X^8,X^9$ directions:
\begin{equation}
\label{chain1}
\text{I}\equiv \frac {\text{IIB} } \Omega   
\xlongrightarrow{T_9}
\text{I}'
\equiv 
 \frac {\text{IIA} }{ \Omega R_9} \xlongrightarrow{T_8} \frac {\text{IIB}}{\Omega R_{89}(-1)^{F_L}}  \xlongrightarrow{T_7}  \frac {\text{IIA}}{\Omega R_{789}(-1)^{F_L}} 
 \xlongrightarrow{T_6}
  \frac {\text{IIB}}{\Omega R_{6789} }.
\end{equation}
Here $R_{ij\dots k}$ denotes a reflection in the directions $ {X^i, X^j\dots X^k}$.
After $p\le 4$ T-dualities, a $p$-torus $T^p$ is identified under reflections so that
the $T^4$ becomes $T^{4-p} \times (T^p/ {\mathbb{Z}_2})$ where $T^p/ {\mathbb{Z}_2}=
T^p/R_{i_1\dots i_p}$ is  identified under the reflection $R_{i_1\dots i_p}$. This has $2^p$ fixed points with an O$(9-p) $-plane at each fixed point of charge $-16/2^p$, which is cancelled by 16 D($9-p$)-branes.

The next case    {IIB}$/{\Omega R_{6789} }$ is an orientifold of type IIB compactified on $T^4/ {\mathbb{Z}_2}$, which has 16 O$5$-planes at the 16 fixed points, each of charge $-1$, together with 16 D$5$-branes to cancel the charge.
Acting with S-duality takes \cite{Sen:1998ii}
\begin{equation}
S:\Omega \to  (-1)^{F_L}
\end{equation}
giving the orbifold $   \text{IIB}/{(-1)^{F_L} R_{6789} }$. The 16 D5-branes become 16 NS5-branes and the 16 O5-planes become 16 ON-planes \cite{Sen:1998ii,Hanany:1999sj}.

Finally, acting with a T-duality   in the $X^6$ direction gives the supersymmetric orbifold 
 $   \text{IIA}/{  R_{6789} }$,
 resulting in
   the IIA string compactified on the K3 orbifold $T^4/ {\mathbb{Z}_2}$:
\begin{equation}
\label{chain2}
  \frac {\text{IIB}}{\Omega R_{6789} }
  \xlongrightarrow{S}
  \frac {\text{IIB}}{(-1)^{F_L} R_{6789} }
\xlongrightarrow{T_6}
 \frac {\text{IIA}}{  R_{6789} }.
\end{equation}
Combining (\ref{chain1}) and (\ref{chain2}) we have the duality between the type I string compactified on $T^4$ and the type IIA string compactified on the K3 orbifold $T^4/ {\mathbb{Z}_2}$. As the type I string is dual to the heterotic string, this gives the duality between the heterotic string on $T^4$ and the type IIA string on K3 \cite{Hull:1994ys}.
Each orbifold singularity of $T^4/ {\mathbb{Z}_2}$ can be resolved by glueing in an Eguchi-Hansen space, and the 16 NS5-branes of the    {IIB}$/{(-1)^{F_L} R_{6789} }$ theory can be thought of as dual to the ALE spaces glued in to resolve the singularities.
Further T-duals will be discussed in sections  \ref{modsp} and
\ref{ngdu}.

\subsection{Moduli Spaces and Dualities}

In  the orientifolds considered in the previous subsection, the total charge cancels between branes and orientifold planes and in general the  dilaton is non-constant. However, in each case there is a particular configuration where the charge cancels locally and the dilaton is constant.
In the type I$'$ theory there are two orientifold planes of charge $-8$ and 16 D8-branes of charge $+1$, so that if  there are $8$ D8-branes at each orientifold plane, the  total  charge at each plane is zero and  the dilaton is constant.
This configuration has an $SO(16)\times SO(16)$ gauge symmetry perturbatively,  which can be enhanced to $E_8\times E_8$ in the non-perturbative theory \cite{Horava:1995qa,Bachas:1997kn,Horava:1996ma}.

In this case, the function $V$ is a constant, $V=c_1$, and the length of the interval with respect to the metric (\ref{metrictypeI'}) is $L=\pi (c_1)^{1/4}$.

For the theory obtained from the type I string on $T^p$ ($p\le 4$) by performing a   T-duality on each of the $p$ circles, the locally charge-cancelling configuration has $16/2^p$ concident D$(9-p)$-branes at each of the $2^p$ O$(9-p)$ orientifold planes. For $p=4$, there   is a single  D5-brane at each of the 16 fixed points, and S-dualising gives  a single NS5-brane coincident with each of the $16$  ON-planes.

The general configuration for each case is obtained by moving in the corresponding moduli space. The type I$'$ string is defined on $I\times \mathbb{R}^{1,8}$ where $I$ is the interval $0\le \tau\le \pi$ with coordinate $\tau$, with 16 D8-branes at positions $\tau_1,\dots , \tau_{16}$ and O8-planes at the end-points $\tau=0,\pi$.
It has an 18-dimensional moduli space
\begin{equation}
O(1,17;\mathbb{Z}) \backslash   {O(1,17) }/{O(17)} \times \mathbb{R}^+
\end{equation}
which includes the coset space ${O(1,17) }/{O(17)}$ identified under the action of the discrete duality group $O(1,17;\mathbb{Z})$.
The $18$ moduli can be viewed as consisting of the $16$ D8-brane positions $\tau_1,\dots , \tau_{16}$,  the length of the interval $L$
and the dilaton zero-mode.
The duality group $O(1,17;\mathbb{Z})$ (corresponding to the T-duality symmetry of the heterotic string compactified on $S^1$) acts on all 17 of the coset moduli.

For the theory obtained from the type I string on $T^p$ ($p\le 4$) by performing a   T-duality on each of the $p$ circles, the moduli space
is 
\begin{equation}
O(p,16+p;\mathbb{Z}) \backslash   {O(p,16+p) }/{O(p)\times O(16+p)} \times \mathbb{R}^+
.
\end{equation}
The moduli consist of the $16p$ parameters that can be interpreted as determining the positions of the 16  D$(9-p)$-branes on $T^p/\mathbb{Z}_2$, the $p^2$ moduli of constant metrics and RR 2-form gauge-fields on $T^p/\mathbb{Z}_2$ and the dilaton zero-mode. The moduli space for metrics and 2-form gauge-fields on $T^p$
  is
\begin{equation}
O(p,p;\mathbb{Z}) \backslash
  {O(p,p) }/{O(p)\times O(p)}.
\end{equation}

For $p=4$, we obtain \begin{equation}
\label {K3mod}
O(4,20;\mathbb{Z}) 
\backslash
  {O(4,20) }/{O(4)\times O(20)} \times \mathbb{R}^+
.
\end{equation}
After S-duality, this becomes the 81-dimensional moduli space consisting of $4\times 16$ positions of NS5-branes, 16 moduli for metrics and NS-NS 2-form gauge fields on $T^4/\mathbb{Z}_2$ and the dilaton zero-mode.
An $O(4,4;\mathbb{Z})$ subgroup of $O(4,20;\mathbb{Z})$ acts as T-duality on $T^4$, while the remaining transformations mix the NS5-brane positions with torus moduli.

For IIA on K3, the moduli space is again (\ref{K3mod})~\cite{Aspinwall:1996mn}; here 
  $O(4,20;\mathbb{Z})$  is the automorphism group    of the K3 CFT and  contains large diffeomorphisms, shifts of the $B$-field  and mirror transformations. Moving in the moduli space away from the orbifold point blows up the singularities, and generic points in the moduli space correspond to smooth K3 manifolds. 

\section{Dual Configurations}

We will start with the type I$'$  string compactified on $T^3$. With the charge-cancelling configuration of 8 D8-branes coincident with each O8-plane, the geometry is
$ I \times T^3\times \mathbb{R}^{1,5} $ (where $I$ is the interval $[0,\pi ]$) with constant dilaton. For general positions of the D8-branes, the corresponding supergravity solution is (\ref{metrictypeI'}), (\ref{phitypeI'}), (\ref{FtypeI'}) with $x,y,z$ periodically identified.

Applying  T-dualities in the $x,y,z$  directions,
the  supergravity solution transforms according to the
Buscher rules,  
taking each D8-brane to a D5-brane that is smeared over the $T^3$. In other words, instead of getting a D5-brane localised at a point on the 4-dimensional transverse space with a harmonic function $V(\tau,x,y,z)$, we get a harmonic function $V(\tau )$ depending only on $\tau$. 
Applying the standard Buscher T-duality rules takes the solution (\ref{metrictypeI'}), (\ref{phitypeI'}), (\ref{FtypeI'})  to the solution (\ref{metricIIB}), (\ref{FIIB}), (\ref{phiIIB}). This
  suggests that the O8-planes could behave like negative tension D8-branes under T-duality, transforming to O5-planes smeared over the $T^3$.
However, this picture is too naive:
under T-duality, the transverse space $T^3\times S^1/\mathbb{Z}_2$ transforms not to the product of the dual $T^3$ with $ S^1/\mathbb{Z}_2$ but to $T^4 /\mathbb{Z}_2$.

The quotient $S^1/\mathbb{Z}_2$ is the circle with coordinate $\tau \sim \tau + 2\pi$ identified under the action of the $\mathbb{Z}_2$ acting as a reflection $\tau \to - \tau$, so that $\tau=0$ and $\tau = \pi$ are fixed points.
It can be represented by the line interval $I$ with $0 \le \tau \le \pi$.

The orbifold $T^4 /\mathbb{Z}_2$ can be realised in a similar fashion.
The $T^4$ has 4 periodic coordinates $x^\mu= (\tau,x,y,z)$ each identified with 
$x^\mu \sim x^\mu+2\pi$, so that we can take $0 \le x^\mu \le 2 \pi$.
Then this is identified under the reflection acting as $x^\mu \to - x^\mu$.
As each point with $\pi < \tau \le 2 \pi$ is identified with a point with $0 \le \tau \le \pi$, each point in $T^4 /\mathbb{Z}_2$ can be represented by a point in
  $I\times T^3$ where $I$ is the interval with
$0 \le \tau \le \pi$ and the $T^3 $ has periodic coordinates $x,y,z$ each with period $2\pi$.
There is then a further quotient at the end points of $I$, so that $I\times T^3$ is identified under the $\mathbb{Z}_2$ acting to take 
\begin{eqnarray}
(0,x,y,z) \to (0,-x,-y,-z), \qquad (\pi,x,y,z) \to (\pi,-x,-y,-z).
\end{eqnarray}
Thus we have $T^3$ \lq fibred' over $I$ with the fibre a $T^3$ for generic points    $\tau$ with  $0 < \tau < \pi$, but at the end points $\tau = 0, \pi$, the fibre degenerates to $T^3/\mathbb{Z}_2$ with $8$ fixed points on each $T^3/\mathbb{Z}_2$.
If we take the length of the interval $L$ to be large, then the naive supergravity solution can be a good approximation a long way away from the end points, but will need to be modified near $\tau =0,\pi$.

Dualising the D8-brane configuration then gives a configuration that, for $| \tau -\pi/2 | << \pi/2$, is well approximated by the supergravity solution (\ref{NS1}), (\ref{NS3}), (\ref{NS2}) consisting of the three-torus with flux fibred over a line, but this will need modification near the end points $\tau =0,\pi$.
This then is a space with a long neck of the form $T^3\times \mathbb{R}$ capped at the two ends.
A further T-duality takes the fibres from a 3-torus with flux to a nilfold ${\cal N}$, giving the solution (\ref{nilfold1}), (\ref{nilfold2}). This then implies that K3 should have a limit in which it degenerates to a long neck of the form ${\cal N} \times \mathbb{R}$ capped off by suitable smooth geometries.
Remarkably, such a limit of K3 has recently been found \cite{2018arXiv180709367H}, with an explicit understanding  of the geometries needed to cap off the neck and to resolve the domain wall singularities, as we discuss in the next section.
\section{A Degeneration of K3}
\label{degK3}

In \cite{2018arXiv180709367H}, a family of hyperk\"ahler  metrics 
on  K3, labelled by a parameter $\beta$,  was constructed  in which
the limit $\beta \rightarrow \infty$ gives a boundary of the K3 moduli space in which the K3 
 collapses to the one-dimensional line segment $[0,\pi]$. For large $\beta$, the  metric is given to a good approximation at generic points by the multi-domain wall metric (\ref{nilfold1}).
 The domain wall solution $\hat {\cal N}$  (\ref{nilfold1}) has singularities at the end points $0,\pi$ where there are single-sided domain walls and at the domain wall locations $\tau_i$, 
 but the K3 metric of \cite{2018arXiv180709367H}
    resolves these singularities to give a smooth geometry. 
    
    The smooth K3 geometry is obtained by glueing together a number of hyperk\" ahler spaces, and these then give approximate metrics for 
different regions of K3.  
There is a long neck consisting of a number of segments, with each segment a product of  a nilfold  with a line interval, with metric of the form (\ref{nilfold1}) with $V=c+m\tau$. The degree $m$ of the nilfold jumps between segments, and the constant $c$ is chosen in each segment so that $V$ is continuous.   The  domain wall connecting two segments is realised as 
 a smooth Gibbons-Hawking space corresponding to multiple Kaluza-Klein monopoles.
At either end  the geometry is capped    with a Tian-Yau space, resolving the single-sided domain wall geometry, as discussed in section \ref{TYsec}.
 
For  large $\beta \gg 1$, there exists a continuous surjective map from K3 to the interval $I$
\begin{equation}
F_\beta:K3 \rightarrow [0,\pi]
\end{equation}
and a discrete set of points ${\cal S}= \{ 0, \tau_1, \dots, \tau _n, \pi \} \subset [0,\pi]$.  It will be convenient to let $\tau_0=0$ and $\tau_{n+1}=\pi$.
Let ${\cal R}_\epsilon^i$ be the interval
\begin{equation}
{\cal R}_\epsilon^i = \{ \tau:  \tau_{i-1} +\epsilon< \tau < \tau_i -\epsilon \}= ( \tau_{i-1} +\epsilon,  \tau_i -\epsilon 
)
\end{equation}
for some  small $\epsilon$.
Then for each $i=1,\dots , n+1$,  the region $F_\beta^{-1}({\cal R}_\epsilon ^i )$ of K3   projecting to ${\cal R}_\epsilon^i$
is diffeomorphic to the product of  the interval  ${\cal R}_\epsilon^i$ with
a nilfold of degree $m_i$ for some $m_i$.
The metric is approximately given by the hyperk\" ahler metric (\ref{nilfold1}) with
$V=c_i + m_i \tau$.
 The degree of the nilfold fibres is piecewise constant:
 \begin{equation}
M(\tau) = \begin{cases}
m_1, & \tau < \tau_1 -\epsilon\\
m_2, & \tau_1 +\epsilon< \tau < \tau_2 -\epsilon\\
\vdots & \\
m_n,& \tau_{n-1} +\epsilon< \tau < \tau_n -\epsilon\\
m_{n+1},& \tau > \tau_n+\epsilon 
\end{cases}\label{Mep}
\end{equation}
and jumps across the \lq domain walls' at $\tau = \tau_1,\tau_2,\ldots,\tau_n$, with
the degree jumping at $\tau_i$ by
\begin{equation}
N_{i}= m_{i+1}-m_i.
\end{equation}

For the end regions $F_\beta^{-1}({\cal S}_\epsilon ^-)$, $F_\beta^{-1}({\cal S}_\epsilon ^+)$ projecting to
\begin{equation}
{\cal S}_\epsilon ^ -=[0,\epsilon )
\qquad
{\cal S}_\epsilon ^+=(\pi- \epsilon ,\pi]
\end{equation}
 the singularities of the single-sided domain walls at the end points are resolved, as in section  \ref{TYsec}, by Tian-Yau spaces.
 The region  $F_\beta^{-1}({\cal S}_\epsilon ^+)$ is approximately a Tian-Yau space  $X_{b_+}$ of degree $b_+$ and the
 region $F_\beta^{-1}({\cal S}_\epsilon ^-)$ is approximately a Tian-Yau space  $X_{b_-}$ of negative degree $-b_-$, where 
 $b_\pm$ are some integers $0\le b_\pm \le 9$.
 These are each asymptotic to the product of a line with a nilfold, one  of degree $-b_-$ and one of degree
 $b_+$, and so to match with the 
 solutions projecting to ${\cal R}_\epsilon^1$, ${\cal R}_\epsilon^{n+1}$
 we   take
 $m_1=-b_-$ and $m_{n+1}= b_+$.
   Then the sum of the charges 
   is
 \begin{equation}
\sum _{i=1}^{n} N_i= b_-+b_+
\label{sumN}
\end{equation}
and so is an integer in the range
\begin{equation}
0\le \sum _{i=1}^{n} N_i \le 18.
\end{equation}
The number $n$ of domain walls then satisfies $0\le n\le 18$ if all  domain wall charges $N_i$ are positive.

 Consider now the interval
 \begin{equation}
{\cal S}_\epsilon ^i =( \tau_i -\epsilon , \tau_i +\epsilon ).
\end{equation}
In  \cite{2018arXiv180709367H}, the geometry in the region 
$F_\beta^{-1}({\cal S}_\epsilon ^i )$ is taken to be approximately a Gibbons-Hawking metric (\ref{metricGH})
specified by a harmonic function $V(x,z,\tau)$ on 
$T^2\times {\cal S}_\epsilon ^i$ with $N_i$ sources. This gives a hyperk\" ahler space which is an $S^1$ fibration over 
the space given by removing the $N_i$   points from 
$T^2\times {\cal S}_\epsilon ^i$.
This can be constructed from a Gibbons-Hawking space on  $\mathbb{R}^2\times {\cal S}_\epsilon ^i$ with a doubly periodic array of sources, 
with contributions summed as in~\cite{Ooguri:1996me},
by taking the quotient by a lattice to obtain
 $N_i$ sources on $T^2\times {\cal S}_\epsilon ^i$. 
 There is a \lq bubbling limit' in which this region is mapped to a  Gibbons-Hawking space  \cite{2018arXiv180709367H}, making precise the sense in which the Gibbons-Hawking space is an approximation.
 The neck region is then said to have Gibbons-Hawking or KK monopole bubbles.
 
 The result of \cite{2018arXiv180709367H} is that these hyperk\" ahler spaces can be glued together to give a  K3 manifold with a smooth hyperk\" ahler metric. The model geometries -- the Tian-Yau spaces, the nilfold fibred over a line and the Kaluza-Klein monopole spaces -- then provide good approximate metrics in each of the corresponding regions.

For the case $b_-=b_+=0$, both $X_{b_+}$ and $X_{b_-}$ are asymptotically cylindrical ALH spaces, with fibres given by $T^3$.
A K3 surface can be constructed by glueing  the two  cylindrical ends of  two ALH space together  
with a long cylindrical neck region of the form $\mathbb{R}\times T^3$
 \cite{2016arXiv160308465C}. 
This means there are no domain walls, so that $n=0$ for this case.
This K3 surface is dual to the locally charge-cancelling  type I$'$ configuration, in which one end  of the interval at $\tau = 0$ there is an O8-plane and 8 D8-branes, while at  the other end, $\tau = \pi$, there is also an O8-plane and 8 D8-branes, so that the RR field strength $F_{(0)}$ is zero.

In the general case we have a geometry capped by two  spaces $X_{b_-}$ and $X_{b_+}$ of degrees $b_-,b_+$ which are integers with $0\le b_\pm\le 9$. For $b_\pm >0$ these are 
 Tian-Yau spaces asymptotic to the product of a nilfold of degree $b_\pm$ and a line interval, while for $b_\pm =0$ these are ALH spaces asymptotic
 to the product of a 3-torus and a line interval. These are joined by a neck region which can be thought of as a
 Gibbons-Hawking space on the product of a line interval and a nilfold with  $b_{-}+b_{+}$ Kaluza-Klein monopoles inserted.
For $N_i$  Kaluza-Klein monopole bubbles  at $\tau=\tau_i$, the degree of the nilfold jumps from
$m_i$ for $\tau<\tau_i$ to $m_i+N_i$ for $\tau>\tau_i$.
The smooth K3 geometry is constructed by glueing together 
the
Tian-Yau spaces, the product of the nilfold with a line interval and the    Gibbons-Hawking spaces as shown in \cite{2018arXiv180709367H}, and these various hyperk\" ahler metrics provide good approximate metrics for the corresponding regions of the  K3.

The form of the solution away from the domain walls is (\ref{nilfold1}) with $V$ given by (\ref{Vtau1taun})
with
\begin{equation}
m_1= -b_-, \qquad
m_{n+1}= b_+ ,
\qquad
N_{i}= m_{i+1}-m_i
\end{equation}
and the charges $N_i$ satisfy (\ref{sumN}).

The geometry is smooth if all   of  the Kaluza-Klein monopoles are at distinct locations, so that the geometry is approximately that of self-dual Taub-NUT near each monopole.
If $k$ of the Kaluza-Klein monopoles are coincident, the K3 surface  has an $A_{k-1}$ orbifold singularity and  there is a resulting $A_{k-1}$ gauge symmetry.
With $b_-=b_+=9$, there are $18$  Kaluza-Klein monopoles and if these are coincident, there is a resulting $SU(18) $ gauge symmetry.

A set of $n$ D8 branes coincident with an  O8-plane has charge $n-8$ and is dual to a Tian-Yau space of degree $n-8$. Moving a further D8 brane to the O8 plane increases the charge to $n-7$ and corresponds in moving a Kaluza-Klein monopole to the Tian-Yau space, changing the degree to $n-7$. This corresponds to blowing up a point of the original del Pezzo surface.

In the following sections, the duals of the IIA string compactified on this degenerate K3 will be considered.

\section{Matching Dual Moduli Spaces}
\label{modsp}

In this section, we revisit the chain of dualities discussed in section \ref{dualsect} that led from the type I$'$ string theory to the IIA string on K3.
Starting from the local charge-cancelling configuration of the type I$'$ string on $I\times \mathbb{R}^{1,8}$ with 8 D8-branes at each O8-plane, dualising took us to
the type IIA string on the K3 orbifold $T^4/ \mathbb{Z}_2$ or to the 
quotient  $   \text{IIB}/{(-1)^{F_L} R_{6789} }$ of type IIB compactified on $T^4/ {\mathbb{Z}_2}$.
The equivalence of these theories at one point in moduli space then, in principle, should give an embedding of the moduli space $O(1,17;\mathbb{Z}) \backslash   {O(1,17) }/{O(17)}$ of the type I$'$ string theory into the moduli space of the
IIA string on K3 and of the IIB quotient, and give an equivalence between the theories for all points in
$O(1,17;\mathbb{Z}) \backslash   {O(1,17) }/{O(17)}$, so that  moving in the  type I$'$ moduli space is dual to a corresponding movement in the  moduli space of type II  on K3.

The domain wall supergravity solutions provide a guide as to how this should work.
Moving in the moduli space of the type I$'$ string moves the $16$ D8-branes away from the O8-planes to generic points $\tau_i$ on the interval, corresponding to the solution (\ref{metrictypeI'}) for generic points  away from the locations of the branes. 
Dualising   takes the solution (\ref{metrictypeI'}) with D8-branes  to the solution (\ref{NS1}) with smeared NS5-branes and a $T^3$ fibration over a line or to the solution (\ref{nilfold1}) of smeared KK-monopoles
with a nilfold fibration over a line.
The locations of domain walls arising from the smeared  NS5-branes or KK-monopoles are at the same locations $\tau _i$.
The geometry discussed in the previous section then provides a non-singular K3 geometry that resolves the singularities of the domain-wall supergravity solution, and its existence supports the picture arising from duality arguments.

The type I$'$ configuration with two O8-planes of charge $-8$ and 16 D8-branes then corresponds to the K3 geometry with end-caps given by Tian-Yau spaces with 
$b_+=b_-=8$ and with  16 Kaluza-Klein monopoles distributed over the interval.
If $b_+=8-n_+$ and $b_-=8-n_-$ with $16-b_--b_+$ Kaluza-Klein monopoles,
 this corresponds in the type I$'$ string  to   having $n_-$ D8-branes at the O8-plane at $\tau=0$ and 
$n_+$ D8-branes at the O8-plane at $\tau=\pi$, with $16-b_--b_+$ D8-branes distributed over the interval.

However, the K3 geometry also allows $b_+=9$ and/or $b_-=9$, which would lead to up to 17 or 18 Kaluza-Klein monopoles.
This corresponds to the possibility in the type I$'$ string at  strong coupling for an O8-plane to emit a D8-brane leaving an O$8^*$-plane of charge $-9$ \cite{Morrison:1996xf,Gorbatov:2001pw}.
Then the K3 with $b_+=b_-=9$ and $18$ Kaluza-Klein monopoles corresponds in the type I$'$ string to  two  O$8^*$-planes of charge $-9$ and 18 D8-branes.
The configuration in which the 18 Kaluza-Klein monopoles are coincident corresponds to the one in which the  18 D8-branes are coincident, and either picture gives an enhanced gauge group $SU(18)$ (together with a further $U(1)$ factor).
For the type I$'$ string, up to 16 D8-branes are possible at weak coupling and 17 or 18 D8-branes are only possible at strong coupling.
However, for the K3 geometries, the IIA string theory on K3 can be taken at weak IIA string  coupling and  in particular 17 or 18 KK-monopoles and the gauge group $SU(18)$ can arise at weak coupling.
The S-duality in the chain of dualities in section  \ref{dualsect} has mapped strong coupling physics  of the type I$'$ string to weak coupling physics in the
IIA string on K3.

For degree $b=8$, there are two distinct Tian-Yau spaces, corresponding to the two distinct del Pezzo surfaces of degree 8, 
 $\mathbb{CP}^1\times\mathbb{CP}^1$ and   $\mathbb{CP}^2$ with one point
 blown up.
One of these  Tian-Yau spaces is presumably dual to the standard O8 plane with charge $-8$, while the other is presumably dual to a variant of this, also with charge $-8$, and it would be interesting to understand this further; in particular, it would be interesting to understand whether there is a relation to the variant O8 planes  discussed in e.g. \cite{Hanany:2000fq,Bergman:2013ala}.
In particular, it seems that the  IIA theory at weak coupling  is revealing some interesting structure in the strongly coupled dual type I$'$ theory.

Consider now the type IIB dual of the K3 compactification.
Compactifying the weakly-coupled  I$'$ string on $T^3$ and dualising on all three toroidal directions gave 16 D5-branes smeared over the  $T^3$  and 16 O5-planes.
The IIB theory is an orientifold on $T^4/\mathbb{Z}_2$, which can be regarded as $I\times T^3$ with an identification of the  3-tori at the ends $\tau =0,\pi$  of the interval to become $T^3/\mathbb{Z}_2$, with an O5-plane at each fixed point.
S-dualising gives the quotient    $   \text{IIB}/{(-1)^{F_L} R_{6789} }$
with 16 ON-planes and 16 NS5-branes smeared over the $T^3$.
This should be dual to the IIA string  on K3, and for the    K3 orbifold $T^4/\mathbb{Z}_2$ they  are related by  a T-duality.
However,  the relation between the IIA and IIB pictures cannot be a conventional T-duality at generic points in the moduli space, as Buscher  T-duality requires the geometry to have an isometry and a smooth  K3 does not have any isometries.

The degenerating K3 geometry of~\cite{2018arXiv180709367H} is constructed by glueing a number of hyperk\" ahler segments.
The  segment with geometry $\hat {\cal N}$ (\ref{nilfold1}) with a nilfold fibred over a line segment dualises to (\ref{NS1}) with a 3-torus with flux fibred over  a line  segment.
The Tian-Yau caps do not have the required isometries and so do not have conventional T-duals. However, from the duality with the type I$'$ theory, they should be dual 
to the region around the $8$ ON-planes. The Tian-Yau caps are asymptotic to the nilfold fibred over a line, and so their duals should be asymptotic to a $T^3$ with flux fibred over  a line.

The segment near the domain wall of charge $N_i$ at $\tau=\tau_i$ is realised in the K3 geometry as $N_i$ Kaluza-Klein monopoles on ${\cal N}\times I$, arising as a Gibbons-Hawking metric with $N_i$ sources on the base space $T^2\times I$.
T-dualising on the $S^1$ fibre of this Gibbons-Hawking  space takes the  $N_i$ Kaluza-Klein monopoles on ${\cal N}\times I$ to $N_i$ NS5-branes  on $T^3\times I$.
This can be understood by first looking at the covering space $ {\mathbb{R}}^2\times I$ of $T^2\times I$.
A single Kaluza-Klein monopole in  $ {\mathbb{R}}^3$ is given by (\ref{KKsolution}) in terms of  
the Gibbons-Hawking form of the Taub-NUT metric (\ref{metricGH}) with
$V(\tau,x,z)$ a harmonic function on  $ {\mathbb{R}}^3$ given by
\begin{equation}
V=c+\frac 1 {| {\bf r}-{\bf r}_0|}
\label{harmo}
\end{equation}
with 3-vector  ${\bf r} =(\tau,x,z)$. T-dualising gives the NS5-brane (\ref{NS1}) with harmonic function
$V(\tau,x,y,z)$ on  the transverse $ {\mathbb{R}}^4$ given again by (\ref{harmo}). It is independent of the coordinate $y$ that is T-dual to the Gibbons-Hawking fibre coordinate and so the solution is smeared in the $y$ direction. Taking a periodic array of such solutions
 in the $x,z$ directions and summing as in \cite{Ooguri:1996me}
 allows periodic identification of the $x,z$ coordinates and gives the GH solution 
localised on $T^2\times {\mathbb{R}}$.
 T-dualising in the $y$ direction gives
  the NS5-brane on $T^3\times {\mathbb{R}}$, smeared over one of the $y$ direction.
  For the segment near  $\tau=\tau_i$, one takes a  superposition of $N_i$ sources giving the Gibbons-Hawking solution with $N_i$ sources on $T^2\times I$.

 At the level of supergravity solutions, the  D8-brane domain wall supergravity solutions   wrapped on $T^3$
map to the KK-monopole domain walls with Gibbons-Hawking metric smeared over two transverse directions $(x,z)$, so that $V(\tau,x,z)$ is independent of 
$(x,z)$.
We have seen that these singular domain walls are resolved to give a Gibbons-Hawking metric with local sources at points  $(\tau,x,z)$
in $T^2\times I$.
The smeared NS5-brane domain walls can also  be replaced by  NS5-brane solutions with sources localised at $N_i$ points in the transverse
$T^3\times I$, as in~\cite{Gregory:1997te}.
Then S-dualising to D5-branes, this would lead to the D5-brane domain wall T-dual to a D8-brane of charge $N_i$ realised as $N_i$ localised D5-branes on
$T^3\times I$. 
 T-dualising to D$p$-branes, we would then have the smeared D$p$-branes replaced by  
 $N_i$ local sources on the transverse $T^{8-p}\times I$.


\section{Non-Geometric Duals}
\label{ngdu}

Compactifying the type I string on $T^4$ and T-dualising in all four torus directions 
and then taking the S-dual gives the quotient
  $   \text{IIB}/{(-1)^{F_L} R_{6789} }$ of the IIB string on $T^4/\mathbb{Z}_2$.
This can then be T-dualised in one, two, three or four directions, leading to new string theory configurations.

The first T-duality works well, as has been discussed in the preceding 
sections.
At the locally-charge-cancelling orbifold point, the T-duality takes this to the orbifold   $   \text{IIA}/ R_{6789} $ of  the IIA string on the K3 orbifold $T^4/\mathbb{Z}_2$.
At generic points in the moduli space of configurations dual to the type I$'$ string, this becomes a duality between a IIB configuration of NS5-branes and ON-planes and the IIA string on
a smooth K3 manifold near the boundary of moduli space in which the K3 becomes a long neck capped with Tian-Yau spaces.
As a smooth K3 has no isometries, this duality is not properly a T-duality but instead a dual form of the duality between IIA on K3 and the heterotic string on $T^4$ \cite{Hull:1994ys}
(which is in turn dual to the type I string on $T^4$).
However, in the long neck region, the geometry is well approximated by a Gibbons-Hawking space and the T-dual of this gives the appropriate configuration of NS5-branes, so dualising the corresponding supergravity solutions gives a good guide to how the duality works.

The first T-duality  takes NS5-branes to KK-monopoles, with NS5-branes on $T^3\times I$ mapped to KK-monopoles on
${\cal N} \times I$.
T-dualising in two or more directions takes the NS5-branes to  exotic branes, so will result in string theory in a non-geometric background.
We now explore this in more detail. 

Consider first T-dualising    $   \text{IIB}/{(-1)^{F_L} R_{6789} }$ in two directions, taking an NS5-brane to 
an exotic $5_2^2$-brane or $(5,2^2)$-brane.
In the last section, we have seen how the naive T-duality between the supergravity solutions representing KK-monopole domain walls and NS5-brane domain walls becomes a proper string theory duality between a smooth K3 geometry and an NS5-brane configuration.
A  T-duality in the $z$ direction takes the nilfold (\ref{nilfold}) to the T-fold (\ref{gb}), and takes the solution (\ref{nilfold1}) with the nilfold fibred over a line to the solution (\ref{Tfold1}) with a T-fold fibred over a line.
If this is subsumed into a  proper string theory duality, then 
this  implies that the degenerate K3 solution with a long neck given by the nilfold fibred over a line is dual to a non-geometric configuration with a long neck given by
the  T-fold fibred over a line. 
More precisely, it should consist of segments each consisting of a T-fold fibred over a line with different  charges for the T-fold for each segment.
In the dual to the supergravity solution, the  segments are separated by domain walls that are smeared exotic branes;
these are expected to become localised exotic branes in the full string theory dual.
The neck configuration can be thought of as a T-fold fibred over a line with bubbles or insertions  of exotic branes, just as the K3 neck can be thought of as a nilfold fibred over a line with Kaluza-Klein monopole bubbles or insertions.
The ends of the neck should 
be capped off by
configurations that can be thought of as 
 the duals of the geometries that cap off the degenerate K3 or as the
 double T-dual of the ON-planes on the fixed points of the ends $T^3/\mathbb{Z}_2$ of $T^4/\mathbb{Z}_2\sim I\times T^3$, and these are  presumably non-geometric.

 Similar remarks apply to T-dualising    $   \text{IIB}/{(-1)^{F_L} R_{6789} }$ in three directions, taking an NS5-brane to an
 exotic 
  $(5,3^2)$-brane or $5_2^3$-brane.
 This takes the configuration $T^3\times I$ with $H$-flux to the configuration  with R-flux that is not  geometric even locally. 
 It cannot be formulated as a conventional background but can be formulated as a doubled geometry, with explicit dependence on the coordinates dual to string winding.
 The doubled geometry of this configuration will be discussed elsewhere.


\section{Discussion}

 Strings propagating on a spacetime with a K3 factor have been much studied. However, the absence of an explicit metric for K3 makes some issues hard to analyse,
and for these an approximate metric can be very useful. The construction of~\cite{2018arXiv180709367H} provides a good approximate metric for K3 near a boundary of moduli space at which the space degenerates to a line.
The K3 metric is obtained by glueing together some hyperk\" ahler spaces, and each  hyperk\" ahler metric provides an approximate K3 metric in the relevant region.
The K3 has a long neck which is divided into segments, each of which is approximated by the hyperk\" ahler metric on 
a nilfold fibred over a line. These are joined by regions which are approximated by Kaluza Klein monopoles  -- the Gibbons-Hawking metric arises in a bubbling limit --
and the   ends of the neck are capped with 
spaces which are approximated by
Tian-Yau spaces --   there is  a bubbling limit of the end region that gives a Tian-Yau space.
This background arises in  a  region of the moduli space of type II strings on K3 
which matches the moduli space of the
  dual  
type I$'$ string, with the Kaluza Klein monopole bubbles dual to the D8-branes and the Tian-Yau caps dual to the O8-planes.
This geometric dual to the orientifold planes is interesting and allows strong coupling properties of the O8-plane 
to be addressed at weak coupling in the dual type IIA theory, such as the emergence of O8* planes.

This K3 limit  in turn is dual to a IIB configuration which can be viewed as having  a long neck which is a 3-torus with H-flux fibred over a line with   NS5-brane bubbles and caps that are ON-planes.
Further dualities replace the neck segments with T-folds fibred over a line and replace the Kaluza Klein monopole bubbles with exotic brane bubbles, while leading to exotic duals of the orientifold planes.

 Earlier string theory roles for del Pezzo surfaces have been discussed in
 \cite{Douglas:1996xp,Iqbal:2001ye,HenryLabordere:2002dk,HenryLabordere:2002xh},
 which include suggestions of a mysterious duality involving them.
 Here we have seen a new role for these surfaces as the starting point for the construction of Tian-Yau spaces. These in turn provide geometric duals of orientifold planes, and so the classification of del Pezzo surfaces
provides a  classification of a class of orientifold planes. It would be interesting to see if there can be a link between these various occurrences of del Pezzo surfaces in string theory.

 A central role in this paper has been played by the hyperk\" ahler wall solution that is a nilfold fibred over a line, with $SU(2)$ holonomy.
 Remarkably, there are higher-dimensional analogues of this in which a higher dimensional version of the nilfold is fibred over a line to give a special holonomy space.
 Examples with holonomy $SU(3), SU(4), G_2, Spin(7)$ are given in~\cite{Gibbons:2001ds}.
 In~\cite{Chaemjumrus:2019wrt}, the duals of these spaces are constructed and the extension of the results of this paper to these examples is explored.

\section*{Acknowledgments}

We are grateful to Amihay Hanany and Costas Bachas for helpful discussions. The  work of CH is supported by the EPSRC Programme
Grant EP/K034456/1,
and by the STFC
Consolidated Grant ST/L00044X/1.




\end{document}